\documentclass[conference,twocolumn]{IEEEtran}

\usepackage{xcolor}
\usepackage{soul}
\usepackage{multirow} 
\usepackage{cite}
\usepackage{mathtools}
\usepackage[labelsep=period]{caption}
\usepackage{enumitem}
\usepackage{bm}

\usepackage{caption}
\usepackage{amsmath}
\usepackage{varwidth}
\usepackage{amssymb}
\usepackage{algorithm}
\usepackage{algpseudocode}
\usepackage{mathtools}
\usepackage[cmintegrals]{newtxmath}
\usepackage{array}
\usepackage{fixltx2e}
\usepackage{stfloats}
\usepackage{enumitem}
\usepackage{url}
\usepackage{booktabs,subcaption,amsfonts,dcolumn}
\makeatletter
\newcommand{\distas}[1]{\mathbin{\overset{#1}{\kern\z@\sim}}}%
\newsavebox{\mybox}\newsavebox{\mysim}
\newcommand{\distras}[1]{%
  \savebox{\mybox}{\hbox{\kern3pt$\scriptstyle#1$\kern3pt}}%
  \savebox{\mysim}{\hbox{$\sim$}}%
  \mathbin{\overset{#1}{\kern\z@\resizebox{\wd\mybox}{\ht\mysim}{$\sim$}}}%
}
\makeatother
\ifCLASSINFOpdf
  \usepackage[pdftex]{graphicx}
  \graphicspath{{./SmartGridComm/figs/}}
\else
\fi

\begin{document}

\title{\huge{A Multi-Agent Deep Reinforcement Learning Approach for a Distributed Energy Marketplace in Smart Grids}\vspace*{-0.2cm}}

{\author{\IEEEauthorblockN{Arman Ghasemi, Amin Shojaeighadikolaei, Kailani Jones, \\Morteza Hashemi, Alexandru G. Bardas,  Reza Ahmadi}\IEEEauthorblockA{Department of Electrical Engineering and Computer Science, University of Kansas, Lawrence, KS, USA} 
E-mails: \{arman.ghasemi, amin.shojaei, kailanij, mhashemi, alexbardas, ahmadi\}@ku.edu}}

\maketitle

\vspace*{-0.5cm}

\begin{abstract}
This paper presents a Reinforcement Learning (RL) based energy market for a prosumer dominated 
microgrid.  The proposed market model facilitates a real-time and demand-dependent dynamic 
pricing environment, which reduces grid costs and improves the economic benefits for prosumers. 
Furthermore, this market model enables the grid operator to leverage prosumers' storage capacity 
as a dispatchable asset for grid support applications.
Simulation results based on the Deep Q-Network (DQN) framework demonstrate significant improvements 
of the 24-hour accumulative profit for both prosumers and the grid operator, as well as major reductions in 
grid reserve power utilization.
\vspace*{-0.3cm}

\end{abstract}

%\IEEEpeerreviewmaketitle

%\begin{IEEEkeywords}
%TBD
%\end{IEEEkeywords}

\section{Introduction}
% \vspace{-.05in}
Small-scale power generation and storage technologies, also known as Distributed Energy Resources (DERs),  
are changing the operational landscape of the power grid in a substantial way. Many traditional power consumers 
adopting a DER technology are starting to produce energy, thus morphing from a consumer to a {\it prosumer} (produces and consumes energy)~\cite{site}.  The most common prosumer installations are the residential solar photovoltaic (PV) systems~\cite{aeo2019}.  Although DER integration has the potential to provide multiple benefits to prosumers as well as grid operators~\cite{7781198}, current grid operating strategies fail to leverage DER capabilities at a large scale, mostly due to the lack of modern and intelligent grid control strategies.  
% \mycomment{AB: We mention ``microgrid'' in the abstract and later in the paper but never really define it. This is SGC not a power conference, thus the reviewers may not be familiar with the terminology. This may be helpful: Definition for microgrid (\url{https://nanopdf.com/download/local-grid-definitions_pdf}) Answer : The micro grid is really well known idiom and we can see the terms all over the papers and in power text books, so it does not need to be define}

The residential PV systems likely have excess power generation during peak sun hours which 
usually do not coincide with peak demand hours~\cite{CaliIso}.  
In other words, current residential PV systems are likely to generate excess power during off-peak demand hours when electricity is not a valuable grid commodity, and this excess generation can even contribute to grid instability. Integration of energy storage into 
prosumer setups can potentially rectify this situation by allowing the prosumers to store their excess energy during the peak sun hours and inject it into the grid during the peak demand hours. 
Furthermore, proper coordination and aggregation of this dispatchable prosumers' generation capacity can be leveraged for various grid support services/applications~\cite{8510890 , 8662200} .  

Nevertheless, current popular net-metering compensation schemes do not properly incentivize the prosumers to engage in grid support applications~\cite{NetM}.  The electricity meter in a net-metered household runs backwards when the prosumer injects power into the grid~\cite{NetPV}. At the end of a billing cycle, 
the customer is billed for the ``net'' energy use, i.e.,  the difference between the overall consumed and produced energy, regardless of the actual schedule of injecting energy into the grid. Moreover, prosumers are compensated for the generated electricity at the same fixed retail price irrespective of the time of the day or any grid contingency at hand.  Therefore, there is little incentive for prosumers to engage in any sort of grid support service. 

In this paper, we propose a distributed energy marketplace framework that realizes a real-time, demand-dependent, dynamic pricing environment for prosumers and the grid operator.  The proposed marketplace framework offers a plethora of vital properties to incentivize prosumers' engagement in grid support applications while providing improved economic benefits to prosumers as well as the grid operator, resulting in a ``win-win'' scenario.  The contributions of the framework proposed in this paper can be summarized as follows,
\begin{itemize}[leftmargin=*]
  \item The proposed marketplace framework enables the grid operator to leverage prosumers' storage capacity as a dispatchable asset, while reducing grid cost through offsetting reserve power with prosumer generation. 
  
  \vspace{.03in}
  \item It incentivizes the prosumers to engage in grid support applications by providing higher economic benefits when supporting grid activities. 
  
  \vspace{.03in}
  \item Founded on a reinforcement learning (RL)-based decision-making, our framework handles the high dimensional, non-stationary, and stochastic nature of the problem without the need for abstract explicit modeling and deterministic rules used in traditional approaches. 
  
    \vspace{.03in}
  \item It models prosumers %in a realistic fashion
  with generation, storage capacity, and bidirectional grid injection capability. This yields in a high degree of freedom for cost versus profit optimization and leads to improved overall benefits for all parties.  
\end{itemize}

\begin{figure*}
\centering
    % \makebox[\textwidth][c]{\includegraphics[scale=0.7, trim = 0cm 0cm 0cm 0cm, clip]{SmartGridComm2020-paper/figs/small-scale-microgrid.png}}
    \includegraphics  [width = 7in]{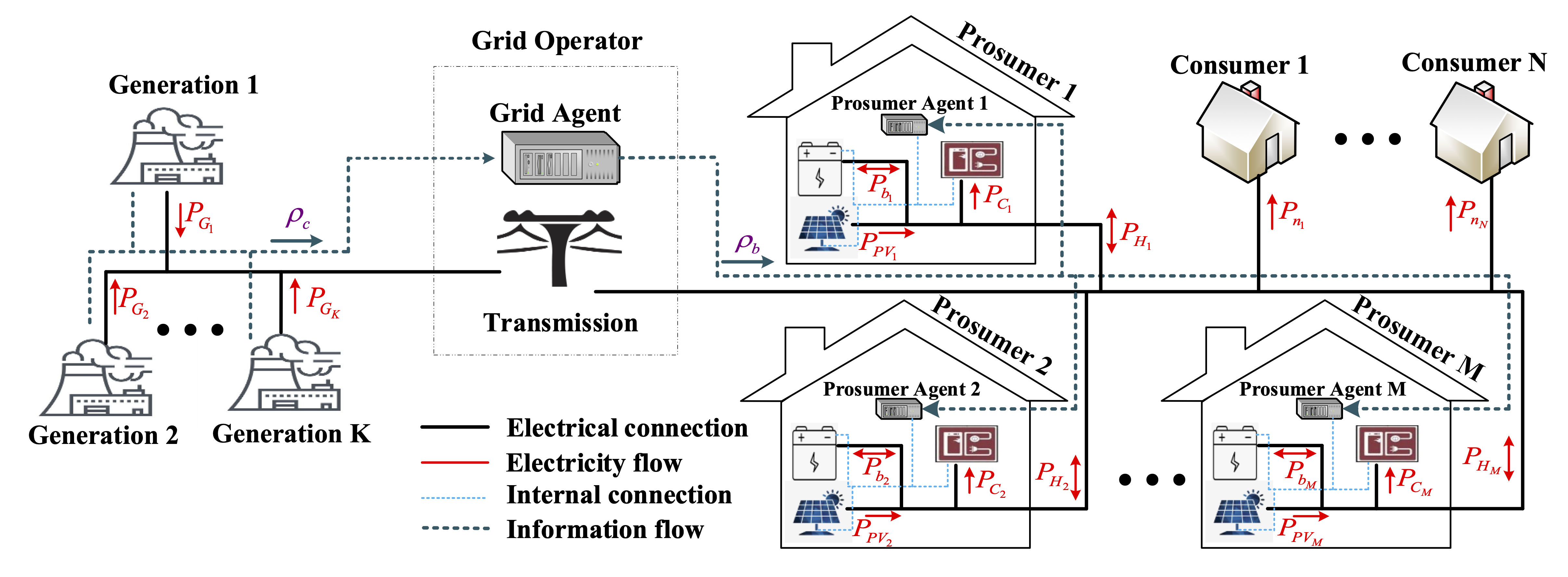}
     \vspace{-.1in}
    \caption{Proposed electricity model market -- The proposed energy marketplace includes several generation sources, household prosumers, and household consumers. By leveraging a reinforcement learning (RL) framework, our system enables a dynamic buy and sell pricing scheme handled by the grid as well as dynamic strategy for the prosumers to maximize benefits.}
    \label{singleLine}
     \vspace{-.1in}
\end{figure*}

To enable all these properties, the proposed energy market leverages a multiagent RL framework with a single grid operator agent, and a network of distributed prosumer agents.  The grid agent's goal is to maximize its economic benefit. To this end, the agent makes decisions on the optimal share of power purchased from a fleet of conventional generation facilities versus a cohort of prosumers with dispatchable generation capability, by considering the incremental cost of generation facilities versus the retail price of purchasing electricity from prosumers.  In order to dispatch the prosumers' generation, the grid agent dynamically sets the retail electricity price to incentivize prosumers to adjust their generation level. On the other hand, the prosumer agents aim to maximize their own economic benefit by deciding on the level of grid support participation according to various factors such as electricity retail price, State of Charge (SoC) of storage device, PV generation level, household consumption level, etc. We demonstrate the efficiency of this marketplace through a simulation on a small scale microgrid as shown in Fig. \ref{singleLine}. The microgrid~\cite{microgrid-definition} is under the management of a single grid operator entity and contains loads, distributed energy resources and/or storage devices that can be operated in a controlled and coordinated way. 

This paper is structured as follows: Section~\ref{sec:background} covers background and related works, while Section~\ref{sec:model} provides the physical and learning system models for the proposed energy market place. Next, the simulation results for the small scale microgrid case study are presented in Section~\ref{sec:simulations}.  Finally,  Section~\ref{sec:conclusions} concludes this paper.    

\section{Background and Related Work}\label{sec:background}
% \vspace{-.05in}
A brief survey of traditional energy marketplace models and dynamic pricing methods for smart grid 
applications is provided in~\cite{8662220 , 7032118, KHAN20161311}.  
On the other hand, research has explored RL-based energy market frameworks and dynamic pricing schemes that bring economic benefits to both costumer and grid operators. The authors in \cite{DynamicPrice} proposed an RL algorithm that allows service providers and customers to learn pricing and energy consumption strategies without a priori of knowledge, leading to reduced system costs. Furthermore, \cite{DynamicEV} investigated an RL-based dynamic pricing scheme for achieving an optimal ``price policy'' in the presence of fast charging electric vehicles over the grid. In order to reduce the electricity bill of the residential customers, a mathematical model using RL for load scheduling was developed in \cite{LoadSchedule}, assuming that residential loads include schedulable loads, non-schedulable loads, and local PV generation. 

More closely aligned to our paper are the works in \cite{11Dynamic} and \cite{12HEM}. \cite{11Dynamic} described an RL-based dynamic pricing, demand response algorithm using Q-learning approach for a hierarchical electricity market that considers both service providers and customers' profits as well as shows improvements in profiability and reduced costs.  However, this work only examines regular customers without generation or storage capacity. The authors in \cite{12HEM} proposed an RL-based home energy management (HEM) framework which considers real-time electricity price and PV generation, and the framework achieve superior performance and cost-effective schedules for demand response in a HEM system. Nonetheless, the households in this work are modeled as traditional loads unable to sell back their excess power to the grid.  Although the Electric Vehicle (EV) charging is modeled, the storage capacity of EVs is not leveraged for cost optimization, meaning the households do not have any energy storage capacity. A demand response dynamic pricing framework is also provided in \cite{NAPS} which is highly related to our work.
 \vspace{-.1in}
\section{System Model}\label{sec:model}
 \vspace{-.1in}
The proposed electricity market model is shown in Fig.~\ref{singleLine}.  As pictured, this model encompasses a grid agent (GA) and several prosumer agents (PAs).  The learning environment is a combination of governing equations of the grid and prosumer’s physical systems, the operational limitations of the power grid and the prosumers, and external factors such as the time of day or PV generation level as explained in the physical model subsection below. Although consumers are depicted in Fig.~\ref{singleLine}, we do not consider them as an individual agent due to their constant consumption of energy. 

\textbf{Notations: } We use the following notations throughout the paper. Bold letters are used for vectors, while non-bold letters are scalars.  Sets are denoted by calligraphy fonts (e.g., $\mathcal{S}$). 
% \blue{In addition, $(.)^H$ denotes the conjugate transpose,} $\text{tr}(.)$ denotes the matrix trace operator, and $\mathds{E}[.]$ denotes the expectation operator. 
The grid and household variables are denoted by $(.)_\text{G}$ and $(.)_\text{H}$.

 \vspace{-0.2in}

% \mycomment{KJ: Our market model doesn't appear to consider the consumer based on the prior sentences. In our simulation, we have one consumer that we assume always consumes energy, and due to this, there is not an agent; however, I would propose that the consumer is still participating in the marketplace and is simply "static" (a.k.a., doesn't inject energy). Due to this, it can be a bit confusing as we end up wrapping the consumer's "action" into the Grid's revenue. What if we mention, with one sentence, why we don't explicitly show the consumer in our equations and why they are not highlighted. Possibly something along these lines, "Although consumers are depicted in Fig.~\ref{singleLine}, we do not consider them as an individual agent due to the constant consumption of energy without injecting energy back into the grid"}.

% The states, actions, and rewards for the agents are also explained further down in the learning system model.

% A variable $X$ for the grid and household is denoted by $X$ and $\tilde{X}$, respectively. 

\subsection{Physical System Model}
% \vspace{-.05in}
\textbf{Grid Operation:} We assume a power system with $K$ generators each with a power output level of $\ P_{G_i}$ such that  $i\in\left\{1,\ldots,K\right\}$, and $M$ prosumers each with power injection level of ${P}_{H_j}$ where $j\in\left\{1,\ldots,M\right\}$.
% and the total grid demand $P_D$,
In the context of an energy marketplace, the goal of the grid is to maximize its profit over a time horizon of $T$, which is denoted by $\psi_\text{G}(T)$.  The accumulative grid profit is then equal to the total grid revenue minus the total cost of operation, i.e., 
% \begin{align}\label{eq:1}
% \textrm{Maximize}{{\ \ \ \psi}_{GA}=\Upsilon_D\left(P_D(t)\right)-  \left\{\sum_{i=1}^{n_g}{\Omega_i\left(P_{G_i}(t)\right)}+  \sum_{j=1}^{n_h}{\Omega_j\left(P_{H_j}(t)\right)}\right\}\ \ },
% \end{align}
\begin{align}\label{eq:1}
 {\psi}_\text{G}(T) = & \Upsilon_\text{G}\left(T \right)-  \left\{\sum_{i=1}^{K}{\Omega_{G_i}(T)}+ \sum_{j=1}^{M}{{\Omega}_{H_j}\left(T\right)}\right\}. 
\end{align}
 In this case, $\Upsilon_\text{G}(\cdot)$ denotes the accumulative grid revenue as a result of selling $P_D(t)$ of electricity to the loads at the selling price of $\rho_s\left(t\right)$ \$/kWh. Therefore, the accumulative revenue over a time horizon of $T$ is defined as:
\begin{align}\label{eq:2}
\Upsilon_\text{G}\left(T\right)= \int_{0}^{T}{P_D\left(t\right){\rho}_s\left(t\right)}dt. 
\end{align}
% , meaning that $\Upsilon_D(P_D(\tau))$ is the accumulative grid revenue  
Moreover, $\Omega_{G_i}(T)$ denotes the accumulative cost of buying electricity from the $i^{th}$ generation facility.  The $\Omega_{G_i}(T)$ is typically estimated using the incremental cost curves of the generation facilities. In addition to the cost of buying electricity from generation facilities, the grid is able to buy electricity from prosumers. Thus, the accumulative cost of buying electricity from the $j^{th}$ prosumer is equal to: 
\begin{align}\label{eq:3}
{\Omega}_{H_j}\left(T\right)=\int_{0}^{T}{{P}_{H_j}\left(t\right)\rho_b\left(t\right)}dt\ \ \ \ \ \textrm{for}\ \ {P}_{H_j}(t)>0\ ,
\end{align}
where  $\rho_b\left(t\right)$  (in the unit of \$/kWh) is the price of purchasing electricity from prosumers, referred to as {\it buy price} hereinafter. 

The GA’s goal is to maximize \eqref{eq:1} subject to the fundamental grid power balance equation,
\begin{align}\label{eq:4}
P_D\left(t\right)-\sum_{i=1}^{K} P_{G_i}\left(t\right)-\sum_{j=1}^{M} {P}_{H_j}\left(t\right)=0\ , \ \  \forall t. 
\end{align}
It should be noted that due to heterogeneous generation facilities, we assume that the output of the $i^{th}$ facility is constrained by practical limitations such as: 
\begin{align}\label{eq:5}
P_{G_i}^{\text{min}}\le\ P_{G_i}(t)\le\ P_{G_i}^{\text{max}}, \ \ \ \text{for} \ i = 1, ..., K. 
\end{align}

\textbf{Prosumer's Operation:} A typical prosumer setup with a PV deployment and energy storage is shown in Fig.\ref{singleLine}.  According to this figure, the goal of the $j^{th}$ prosumer’s agent is to maximize its own accumulative profit ${{\psi}}_{H_j} (T)$ defined as: 
\begin{align}\label{eq:6}
{{\psi}}_{H_j}(T)= {\Upsilon}_{H_j}(T)-{\Omega}_{H_j}(T), 
\end{align}
% $\tilde{\Omega}_j(P_{H_j}(\tau))$ is the accumulative cost of buying electricity from $j^{th}$ prosumer,
where 
% ${\psi}_{H_j}(T)$ is the accumulative profit of the $j^{th}$ prosumer,
${\Upsilon}_{H_j}(T)$ is the accumulative revenue of the $j^{th}$ prosumer for selling electricity to the grid, and ${\Omega}_{H_j}(T)$ is the accumulative cost of buying electricity from the grid defined by:
\begin{align}\label{eq:7}
{\Upsilon}_{H_j}(T)= \int_{0}^{T} {P}_{H_j}(t) \rho_b(t) dt\ \  \ \textrm{for}\ \ {P}_{H_j} (t)>0,
\end{align}
\vspace*{-0.5cm}
\begin{align}\label{eq:8}
{\Omega}_{H_j}(T)=\ \int_{0}^{T}{{P}_{H_j}(t)\rho_s(t)}dt\ \ \ \ \textrm{for}\ \ {P}_{H_j}(t)\le0. 
\end{align} 
Assuming that for the $j^{th}$ prosumer, $P_{PV_j}\left(t\right)$ is the PV generation, ${P_b}_j\left(t\right)$ is battery charge/discharge power, and ${P_C}_j\left(t\right)$ is the consumption power, the internal power balancing is then described as follows: 
\begin{align}\label{eq:9}
{P}_{H_j}\left(t\right)={P}_{PV_j}\left(t\right)- {P}_{b_j}\left(t\right)-P_{C_j}\left(t\right). 
\end{align}
In order to model realistic scenarios, we also pose the following constraints on each of these parameters: 
\begin{enumerate}[label=(\roman*)]
    \item If ${P}_{H_j}^\text{max}$ is the maximum allowable power injection, then we have: $\left|{P}_{H_j}(t)\right|\le\ P_{H_j}^\text{max}$.
    
    \item $P_{{PV}_j}^\text{max}$ denotes the peak PV generation such that $0\le\ P_{PV_j}\left(t\right)\le\ P_{{PV}_j}^\text{max}$. 
    
    \item Given that $P_{b_j}^\text{max}$ is the maximum allowable battery charge/discharge power, then $\left|P_{b_j}\left(t\right)\right|\le\ P_{b_j}^\text{max}$.
    
    \item Assuming that $\phi_j$ is the State of Charge (SoC) of the battery,  and $\phi_j^{\text{min}}$ and $\phi_j^{\text{max}}$ are the minimum and maximum allowable state of charge of battery, we have $\phi_j^{\text{min}}\le\phi_j\le\phi_j^{\text{max}}$.  The state of charge of battery for the $j^{th}$ prosumer is calculated from,
\begin{equation}\label{eq:14}
\phi_j (t)=\phi_j(0)+\frac{1}{C_{B_j}}\int_{0}^{t}{P_{b_j}(\tau)d \tau}, 
\end{equation}
where $C_{B_j}$ is the battery capacity and $\phi_j\left(0\right)$ represents the initial SoC of the battery.
\end{enumerate}
Next we describe a deep reinforcement learning framework to enable the grid and prosumers to dynamically take optimal actions at each time slot.   
%  \begin{align}\label{eq:10}
% \end{align}
% \begin{align}\label{eq:11}
% \ ,
% \end{align}
% \begin{align}\label{eq:12}
% \ ,
% \end{align}
% \begin{align}\label{eq:13}
% \end{align}
% where $P_{PV_j}\left(t\right)$ is the PV generation, ${P_b}_j\left(t\right)$ is battery charge/discharge power, ${P_C}_j\left(t\right)$ is the consumption power, 

\subsection{Reinforcement Learning Model}
In this work, the dynamic pricing problem is formulated as a Markov Decision Process (MDP) such that given a state $s^t$ at time $t$, the goal is choosing the \emph{optimal action} for transitioning to a new state $s^{t+1}$ at time $t+1$, where $s^t, s^{t+1} \in \mathcal{S}$ such that $\mathcal{S}$ is the set of all possible environment states. 
This problem can be viewed as an instance of Reinforcement Learning (RL)
that is concerned with studying how an agent or a group of agents learn(s) the environment by collecting \textbf{observations}, choosing \textbf{actions}, and receiving \textbf{rewards}. Assuming that $\mathcal{A}$
% =\left\{A_{GA},A_{PA_1}\cdots,A_{PA_{n_h}}\right\}$ 
is the set of feasible actions available to each agent, as a result of taking an action $a^t \in \mathcal{A}$, the agent receives an immediate reward $r^t$, and the environment transitions from the state $s^t$ to $s^{t+1}$. 
% P is the set of transition probabilities from any state $s^t\in\ S$ at time slot $t$ to any future state $s^{t+1}\in\ S$ at time slot $t+1$ for any joint action $a^t\in\ A_{GA}\times\cdots\times\ A_{PA_{n_h}}$ at time slot $t$, and $R=\left\{R_{GA}^t,R_{PA_1}^t\cdots,R_{PA_{n_h}}^t\right\}$ is the set of reward functions that determine the immediate reward received by each agent for transition from $\left(s^t,a^t\right)$ to $s^{t+1}$.

In the proposed energy marketplace, we have a set of agents denoted by $\mathcal{N}=\left\{\text{GA}, \text{PA}_1,\ldots,\text{PA}_{M}\right\}$ in which GA is the grid agent and $\text{PA}_j$ is the agent for prosumer $j$. Next, we provide details on the observations, actions, and rewards for each agent type (i.e., grid agent or prosumer agent). In this framework,  all the continuous variables are discretized using a zero-order hold to find the values at each time slot $t$.

% In this case, the MDP framework is defined by tuple $\left(N,S,A,P,R\right)$ where  $S$ is the set of environment states as explained below,

% The environment state vector at time slot $t$ is defined as,
% \begin{align}\label{eq:15}
% {s^t} = \left\{ {\bar \omega _G^t,\bar \omega _H^t,P_D^t,\bar P_H^t,\bar \Phi _H^t,\rho _b^t,\bar P_{PV}^t,\bar P_C^t} \right\} \in S\ ,
% \end{align}

\textbf{\emph{Grid Agent:}} The GA observes the following state variables: 
\begin{enumerate}[label=(\roman*)]
    \item cost of buying electricity from $K$ generation facilities at time $t$, which is denoted by $\pmb{\omega}_G^t=[\omega_1^t, \ldots, \omega_{K}^t]$,  
    \item cost of grid operator for buying electricity from $M$ prosumers, which is denoted by $\pmb{\omega}_H^t=[\omega_{H_1}^t, \ldots, \omega_{H_M}^t]$,  
    \item the total grid demand $P_D^t$,  
    % \item power injections by the $M$ prosumers, which is shown by $\pmb{{P}}_H^t=[{P}_{H_1}^t, \ldots,   {P}_{H_M}^t]$. 
\end{enumerate}
We use the notation ${s}_{GA}^t$ to represent all observations of the grid agent at time $t$. Thus, based on the observations of the grid at time $t$, the grid agent \textbf{action} is to determine the electricity buy price. As described in the physical model, the buy price is denoted by $\rho_b^t \in \mathcal{A}_{GA}$,
% $a_{GA}^t=\rho_b^t\in\ A_{GA}$, 
where $\mathcal{A}_{GA}$ is the finite set of available actions to GA (i.e., all possible buy prices).

% \begin{align}\label{eq:16}
% s_{GA}^t=\left\{\bar{\omega}_G^t,\bar{\omega}_H^t,P_D^t,\bar{P}_H^t\right\}\in\ S\ ,
% \end{align}

The \textbf{reward function for the grid} at time $t$ is defined as the grid profit, i.e., 
\begin{align}\label{eq:18}
r_\text{GA}^t=\upsilon_G^t-\left\{\sum_{i=1}^{K}\omega_{G_i}^t+\sum_{j=1}^{M}{\omega}_{H_j}^t\right\}\ ,
% R_{GA}^t=\psi_{GA}^t=\upsilon\left(P_D^t\right)-\left\{\sum_{i=1}^{n_g}{\omega_i^t\left(P_{G_i}^t\right)+\sum_{j=1}^{n_h}{\omega_j^t\left(P_{H_j}^t\right)}}\right\}\ ,
\end{align}
where $\upsilon_G^t$ denotes the grid revenue at time slot $t$ as a result of selling $P_D^t$ electricity, which is obtained by $\upsilon_G^t=P_D^t{\times\ \rho}_s^t$. In addition, $\omega_{G_i}^t$ is the grid cost to buy $P_{G_i}^t$ from the $i^{th}$ generation facility at time slot $t$. The value of $P_{G_i}^t$ is obtained using incremental cost curve of the $i^{th}$ generation facility. Finally, the grid cost to buy $P_{H_j}^t$ from prosumer $j$ at time slot $t$ is denoted by $\omega_{H_j}^t$ that can be calculated as,
\begin{align}\label{eq:20}
\omega_{H_j}^t ={P}_{H_j}^t\times\rho_b^t\ \ \ \ \ \ \ \ \textrm{for}\ \ {P}^t_{H_j}>0. 
\end{align}
Given the definition for immediate reward $r_{GA}^t$, the ultimate goal is to maximize the agent cumulative reward over an infinite time horizon that is also known as expected return:
\begin{align}\label{eq:24}
 \Gamma_\text{GA}^t=\sum_{k=0}^{\infty}{\gamma^k r_\text{GA}^{t+k+1}}\ ,
\end{align}
where $0\le\gamma\le1$ is the discount rate for the grid agent.

% \vspace{.05in}
\textbf{\emph{Prosumer Agent:}}
The prosumer agent $j$ observes the following state variables: 
\begin{enumerate}[label=(\roman*)]
   \item state of charge of battery that is denoted by $\phi_j^t$, 
   \item PV generation denoted by $P_{PV_j}^t$, 
   \item buy price $\rho_b^t$ determined by the grid agent, 
   \item local power consumption denoted by $P_{C_j}^t$. 
\end{enumerate}

\noindent Based on this set of observations, the charge/discharge command to the energy storage in prosumer $j$ is the \textbf{action} determined by $PA_j$, which is shown by $\sigma_j^t\in\ \mathcal{A}_{PA_j}$. 
% $a_{PA_j}^t=\sigma_j^t\in\ A_{PA_j}$ , 
In this case, $\mathcal{A}_{PA_j}$ is the finite set of available actions to $PA_j$.  
% while $PA_j$ can observe,
% \begin{align}\label{eq:17}
% s_{PA_j}^j=\left\{\Phi_j^t,\rho_b^t,P_{PV_j}^t,P_{C_j}^t\right\}\in\ S\ ,
% \end{align}
%  $\bar{\Phi}_H^t=\left[\Phi_1^t\cdots\Phi_{H_j}^t\right]^T$ is the vector of SoC of battery for $n_h$ prosumers, 
%   $\bar{P}_{PV}^t=\left[P_{PV_1}^t\cdots P_{PV_j}^t\right]^T$ is the vector of PV generation for $n_h$ prosumers, and $\bar{P}_C^t=\left[P_{C_1}^t\cdots P_{C_j}^t\right]^T$ is the vector of local consumption for $n_h$ prosumers.
The \textbf{reward} function for $PA_j$ is defined as,
\begin{align}\label{eq:21}
r_{{PA}_j}^t=\upsilon_{H_j}^t-\omega_{H_j}^t,
\end{align}
where $\upsilon_{H_j}^t={P}_{H_j}^t\times\rho_b^t$ for  ${P}_{H_j}^t>0$ 
is the $j^{th}$ prosumer’s revenue from selling ${P}_{H_j}^t$ to the grid at time slot $t$ and,  $\omega_j^t={P}_{H_j}^t\times\rho_s^t$  for ${P}_{H_j}^t\le0$ 
% \begin{align}\label{eq:23}
% \omega_j^t\left(P_{H_j}^t\right)=P_{H_j}^t\times\rho_s^t\ \ \ \ \ \ \ \ \ \textrm{for}\ \ P_{H_j}\le0\ ,
% \end{align}
is the $j^{th}$ prosumer’s cost from buying ${P}_{H_j}^t$ from the grid at time slot $t$.
% According to the above notations, the MDP trajectories for GA are $s_{GA}^1,\rho_b^1,\psi_{GA}^2,s_{GA}^2,\rho_b^2,\psi_{GA}^3,\ldots$ and its primary goal is to maximize the cumulative reward known as expected return as,
% \begin{align}\label{eq:24}
%  \Gamma_{GA}^t=\sum_{k=0}^{\infty}{(\gamma_{GA})^k\psi_{GA}^{t+k+1}}\ ,
% \end{align}
% where $0\le\gamma_{GA}\le1$ is the discount rate for GA.
% MDP trajectories are $s_{PA_j}^1,\sigma_j^1,\psi_{PA_j}^1,s_{PA_j}^2,\sigma_j^2,\psi_{PA_j}^2,\ldots$ and their 
Similar to the grid agent, the $j^{th}$ prosumer tries to maximize its infinite-horizon accumulative reward defined as:
\begin{align}\label{eq:25}
\Gamma_{{PA_j}}^t=\sum_{k=0}^{\infty}{{\tilde{\gamma}}_j^k r_{{PA_j}}^{t+k+1}}\ ,
\end{align}
where $0\le\tilde{\gamma}_j\le1$ is the discount rate for $PA_j$.

% \vspace{-.01cm}
\subsection{Q-Learning Framework}
In this work, the agents use Deep Q-Network (DQN) to solve their respective MDPs and maximize their accumulative rewards in \eqref{eq:24} and \eqref{eq:25}. The DQN algorithm uses deep learning  for each agent using the bellman iterative equation. In particular, for the grid agent we have, 
% \begin{align}\label{eq:26}
% Q(s_{GA}^t,\rho _b^t) \leftarrow Q(s_{GA}^t,\rho _b^t) + {\alpha _{GA}}[\psi _{GA}^{t + 1} + {\gamma _{GA}}\mathop {\max }\limits_{{\rho ^{t + 1}}} Q(s_{GA}^{t + 1},\rho _b^{t + 1}) - Q(s_{GA}^t,\rho _b^t)]
% \end{align}
\begin{align}\label{eq:26}
Q(s_{GA}^t,\rho _b^t) \leftarrow  Q(s_{GA}^t, &\rho _b^t) +  \nonumber \\ {\alpha}[r_{GA}^{t + 1} + {\gamma}& \mathop  {\max }\limits_{{\rho ^{t + 1}}} Q(s_{GA}^{t + 1},\rho _b^{t + 1}) - Q(s_{GA}^t,\rho _b^t)]\ ,
\end{align}
and similarly, for the prosumer agent we have, 
\begin{align}\label{eq:27}
Q(s_{P{A_j}}^t,\sigma _j^t) \leftarrow  Q( & s_{P{A_j}}^t,  \sigma _j^t) +  \nonumber \\
{\tilde{\alpha} _j}[r_{{PA}_j}^{t + 1} + {\tilde{\gamma} _{j}}& \mathop {\max }\limits_{\sigma _j^{t + 1}} Q(s_{P{A_j}}^{t + 1},\sigma _j^{t + 1}) - Q(s_{P{A_j}}^t,\sigma _j^t)]\ ,
\end{align}
where $\alpha$ and $\tilde{\alpha}_j$ are the learning rates for $GA$ and $PA_j$, respectively. The estimated Q-values are used to find the optimal policy that maximizes the accumulative rewards.  The DQN framework for the grid and prosumer agents is illustrated in Algorithms~\ref{alg1} and \ref{alg2}, respectively. 
\begin{algorithm}[h]
    \caption{Q-learning Algorithm for the Grid Agent}
	\begin{varwidth}{\dimexpr\linewidth-2\fboxsep-2\fboxrule\relax}
		\begin{algorithmic}[1]
			\State Initialize  $Q(s_{GA}^t,\rho_{GA}^t)$ to zero
			\For {each Episode} 
			\For {each Iteration} 
			\State $t: = t + 1$
			\State Set buy price $\rho_b^t$  according to policy $\pi_{GA}$
			\State Observe reward $r_{GA}^{t+1}$  at new state $s_{GA}^{t+1}$
			\State Update $Q(s_{GA}^t,\rho _b^t)$ using \eqref{eq:26}  \State $s_{GA}^t:=s_{GA}^{t+1}$
			\EndFor
			\EndFor
		\end{algorithmic}
		\label{alg1}
	\end{varwidth}%
\end{algorithm}
\vspace*{-0.5cm}
\begin{algorithm}[h]
	\caption{Q-learning Algorithm for the $j^{th}$ Prosumer Agent}
	\begin{varwidth}{\dimexpr\linewidth-2\fboxsep-2\fboxrule\relax}
		\begin{algorithmic}[1]
			\State Initialize $Q(s_{PA_j}^t,\sigma_j^t)$ to zero
			\For {each Episode} 
			\For {each Iteration} 
			\State $t: = t + 1$
			\State Set charge/discharge $\sigma_j^t$   according to policy $\pi_{PA_j}$
			\State Observe reward $r_{PA_j}^{t+1}$  at new state $s_{PA_j}^{t+1}$
			\State Update $Q(s_{PA_j}^t,\sigma _j^t)$ using \eqref{eq:27}  
			\State $s_{PA_j}^t:=s_{PA_j}^{t+1}$
			\EndFor
			\EndFor
		\end{algorithmic}
			\label{alg2}
	\end{varwidth}%
\end{algorithm}

In this framework, to balance exploration versus exploitation, the epsilon greedy strategy $\pi$ is used for GA and PA as follow  \cite{2018IntroductionDeep},
\begin{align*}
\pi  = \left\{ \begin{array}{l}\arg \mathop {\max }\limits_{{a^t}} \,E\left[ {Q\left( {{s^t},{a^t}} \right)} \right]\,\,\,\,\,\,\,\,\,{\rm{with}}\,\,{\rm{probability}}\,\,1 - \varepsilon, \\{\rm{random}}\,\,{\rm{action}}\,\,\,\,\,\,\,\,\,\,\,\,\,\,\,\,\,\,\,\,\,\,\,\,\,\,\,\,\,\,{\rm{with}}\,\,{\rm{probability}}\,\,\,\,\varepsilon. \end{array} \right.\
\end{align*}
The probability of random actions $\varepsilon$ starts at 1 for the first 300 episodes, and then decays to 0.01 over the training episodes.

\section{Case Study and Numerical Results}~\label{sec:simulations}
\begin{figure*}
  \centering
    \includegraphics[scale=0.23, trim =0cm 0cm 0cm 0cm, clip]{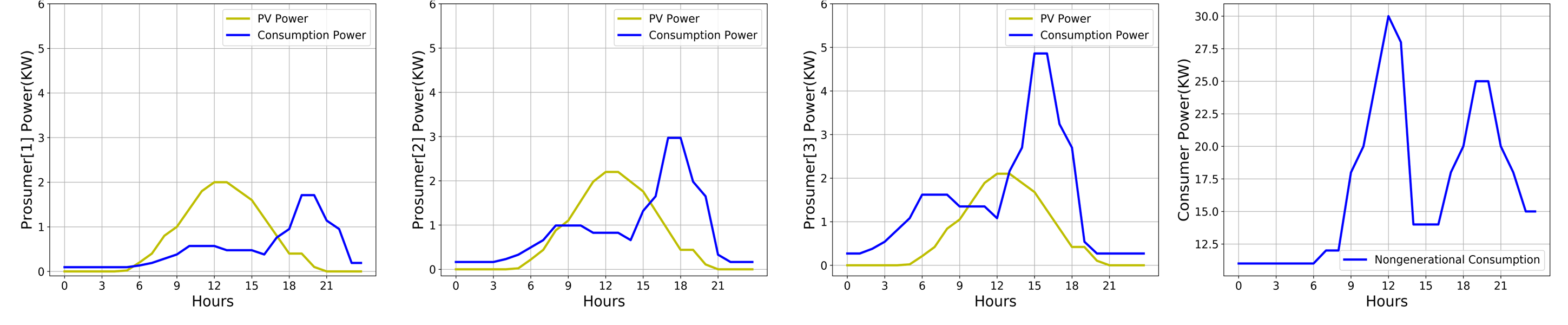}
      \vspace{-.05in}
    \caption{Generation and consumption waveform sample for prosumers and consumer}
    \label{waveform}
     \vspace{-.2in}
\end{figure*}

% \vspace{-.05in}
The proposed energy market place model is implemented on a small-scale microgrid system, illustrated in Fig~\ref{singleLine}, to demonstrate the operation of the agents and their effectiveness for improving the economic benefit of the grid operator and the prosumers. As pictured, the system under the study is comprised of two generation facilities $(K =2)$, three prosumers $(M=3)$ that host the $PA_1$ to $PA_3$ agents, the grid operator that hosts the grid agent (GA), and one nongenerational household (a.k.a., consumer, $N=1$).  The parameters of the system are tabulated in Table~\ref{tab:SimulationParam}. The employed PV generation and local consumption profiles for the last episode of the three prosumers are illustrated in Fig~\ref{waveform}.  These waveforms are constructed to be representative of real-world data available from California ISO website \cite{CaliIso}. The peak value of generation and consumption for each prosumer is listed in Table~\ref{tab:SimulationParam}.  The demand profile for last episode for the nongenerational household is also shown in Fig~\ref{waveform}, and its peak value is listed in Table~\ref{tab:SimulationParam}.  Each prosumer is equipped with an energy storage system (ESS) which includes a constant charge/discharge rate and a capacity provided in Table~\ref{tab:SimulationParam}. 

In order to establish a baseline for the economic benefit of the grid operator and the households, a conventional system with a fixed buy price and no intelligent prosumer agents is simulated.  In this scenario, the prosumers only sell electricity to the grid when their generation is more than their local consumption and their ESS is fully charged, which is likely to happen during the peak sun hours~\cite{economicModel}.  The described microgrid model for trading electricity between grid and residential loads is shown in Fig. \ref{singleLine}. %\textcolor{red}{Fig~\ref{singleLine}} \mycomment{AB: Unclear. Prosumer logic described in Figure 1?}.
This scenario is referred to as the \emph{conventional scenario}.  

\begin{table}[t]
% \scriptsize 
    \resizebox{\columnwidth}{!}{
    \begin{tabular}{l|c|l}
      \hline
      \textbf{Parameter} & \textbf{Description} & \textbf{Value}\\
      \hline
      $P_{p{v_j}}^{\max }$ & Max. PV Generation & [2-2.5] kW \\
      $P_{b_j}^{\max }$ & Max. allowable charge/discharge & 2/-2 kW \\
      $P_{H_j}^{\max }$ & Max. allowable power injection & 10 kW \\
      $\phi_j^{\max }$ & Max. state of charge & $0.9 \times {C_{b_j}}$ \\
      $\phi_j^{\min }$ & Min. state of charge & $0.1 \times {C_{b_j}}$ \\
      $C_{b_j}$ & Energy storage capacity & [8-10] kWh \\
      ${\phi _j}(0)$ & Initial state of charge & [3-4] kWh \\
      $\rho _s$ & Sell price [before 11am, after 11am] & [0.05, 0.095] \$/kWh \\
      $\rho _b^t$ & Buy price for agent-based scenario & ${\{0.05,0.06,0.07,}$ \\ 
                  &                                    & ${0.08,0.09,0.1}\}$\$/kWh \\
      $\rho _b^t$ & Buy price for conventional scenario & 0.05 \$/kWh \\
      $\left[ {P_{{G_1}}^{\min },P_{{G_1}}^{\max }} \right]$ & Limitation of base generation & [5, 20] kW  \\
      $\left[ {P_{{G_2}}^{\min },P_{{G_2}}^{\max }} \right]$ & Limitation of reserve generation & [0, 50] kW  \\
      $\left[ {\beta_1,\beta_2} \right]$ & Incremental cost of two generators & [0.03, 0.3] \$/kWh  \\
      
      \hline
    \end{tabular}
    }
    % \vspace{-.05in}
    \caption{Simulation parameters used for the proposed energy market place model on a small-scale microgrid}
    \label{tab:SimulationParam}
     \vspace{-.3in}
\end{table}

In the next scenario, we leverage the grid and prosumer agents to help implement the proposed market model, and these results are compared with the conventional scenario to demonstrate the economic improvements.  This scenario is referred to as the \emph{agent-based scenario}.  In this work, we use PyTorch framework (v. 1.5.0 with Python3) to implement the DQN agents~\cite{naderializadeh2019energy}.  For training and testing the neural network, we leverage an Intel Xeon processor running at 3 GHz with 16 GB of RAM.The DQN algorithm hyperparameters used for simulations are provided in Table~\ref{tab:DQNhyperparameter}.The simulations for both the conventional and agent-based scenarios are carried out via episodic iterations for 10,000 episodes.  Each episode represents a 24 hour cycle and consists of 96 iterations, meaning that the simulation timeslots are 15 minutes.   

\begin{table}[t]
  \resizebox{\columnwidth}{!}{
    \begin{tabular}{l|c|l}
      \hline
      \textbf{Hyperparameters} & \textbf{Value for $GA$} & \textbf{Value for $PA_j$}\\
      \hline
      Batch size & 64 & 64 \\
      Discount factor & ${\gamma}$=[0.95-0.99] & $\tilde{\gamma}_j$=[0.95-0.99] \\
      Learning rate & $\alpha$=1e-3 & ${\tilde{\alpha}_j}$=1e-3 \\
      Soft update interpolation & 1e-5 & 1e-5 \\
      Hidden Layer-nodes & 1-[1000] & 2-[1000,1000] \\
      Activation & Sigmoid & Sigmoid \\
      Optimizer & Adam & Adam \\
      \hline
    \end{tabular}
    }
    % \vspace{-.05in}
    \caption{DQN hyperparametrs}
    \label{tab:DQNhyperparameter}
     \vspace{-.2in}
\end{table}

The action space for all prosumer agents (i.e., set $\mathcal{A}_{PA}$) includes three options: charge, no charge or discharge, and discharge.  
% $PA_1$ to $PA_3$ is defined as ${A_{P{A_1}}} = {A_{P{A_2}}} = {A_{P{A_3}}} = \left\{ {{\mathop{\rm charge}\nolimits}, \textrm{no\,charge/discharge, discharge}} \right\} $
% which is comprised of three actions to be able to 
As a result, these actions command the battery power to one of the following three levels at each time slot $t$:
\begin{align}
P_{{b_j}}^t = \left\{ \begin{array}{l}P_{{b_j}}^{\max }\,\,\,\,\,\,\,\,\,\,\,{\text{Charge action,}}\\0\,\,\,\,\,\,\,\,\,\,\,\,\,\,\,\,\,\,\,\,{\text{No charge or discharge action,}}\\ - P_{{b_j}}^{\max }\,\,\,\,\,\,\ {\text{Discharge action.}}\end{array} \right.\
\end{align}
The action space for GA (i.e., buy price) is defined as ${\mathcal{A}_{GA}} $= \{0.05, 0.06, 0.07, 0.08, 0.09, 0.1\} in which all numbers represent \$/kWh values.  The sell price $\left( {\rho _s^t} \right) $ is defined at a constant rate in this work as provided in Table~\ref{tab:SimulationParam}.  The incremental cost of the two generators in terms of \$/kWh are defined as,
\begin{align}\label{eq:30}
\left\{ \begin{array}{l}\omega _{G_1}^t = {\beta _1}\,\,\,\,\,\,\,\,\,\,\textrm{for}\,\,\,\,\,\,\,\,\,P_{{G_1}}^{\min }\, \le P_{{G_1}}^t \le P_{{G_1}}^{\max }\\\omega _{G_2}^t = {\beta _2}\,\,\,\,\,\,\,\,\,\textrm{for}\,\,\,\,\,\,\,\,\,P_{{G_2}}^{\min }\, \le P_{{G_2}}^t \le P_{{G_2}}^{\max }\end{array} \right.\ .
\end{align}
where ${\beta _2} > {\beta _1}$ (see Table~\ref{tab:SimulationParam}).  Consequently, the ${P_{{G_1}}}$ provides baseline generation capacity at a lower incremental cost while ${P_{{G_2}}}$  provides reserve capacity at a much higher cost.     

\begin{figure}[t]
  \centering
    \includegraphics[scale=0.22, trim =0cm .5cm 0cm 0cm, clip]{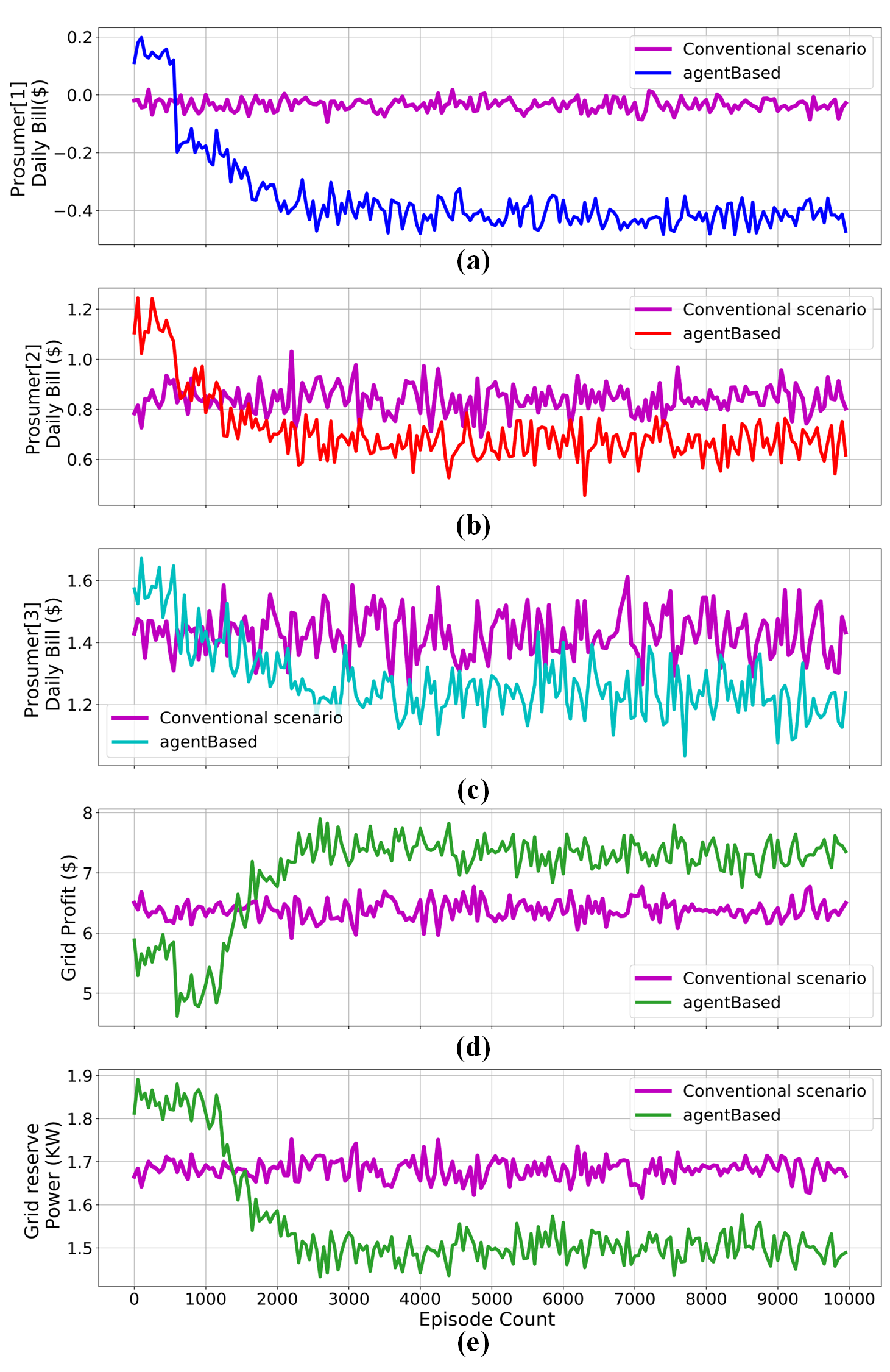}
    \vspace{-.1in}
    \caption{Simulation results for  conventional vs. agent-based scenarios over 10000 episodes:(a)-(c) 24-hour accumulative reward comparison for three prosumers, (d) grid 24-hour accumulative reward comparison, (e) grid reserve power utilization.}
    \label{profit}
     \vspace{-.2in}
\end{figure}

The simulation results comparing the conventional and agent-based scenarios throughout 10,000 episodes are illustrated in Fig.~\ref{profit} (a)-(c), where we compare the daily bill of the three prosumers over a 24-hour period.  From the results, we note that while the daily bill resulting from a conventional scenario remains fairly constant throughout the episodes, the prosumer agents start converging to a lower bill as the agents explore the environment further and learn the optimal policy.  As shown, the daily bill for households 1-3 are lowered by 1400\%, 27\%, and 13\%,  respectively.  The unusually high daily bill reduction for household 1 is attributable to the conventional daily bill that is close to zero since the beginning (i.e., high PV generation), and the household’s smaller peak consumption according to Fig.~\ref{waveform}.  

Fig.~\ref{profit} (d)-(e) compare the accumulative grid profit and use of costly reserve power (PG2) over a 24-hour period. The agent-based scenario starts with a lower profit than the conventional scenario but converges to a much higher profit level than the conventional scenario as the agent learns the optimal policy.  In this case, the grid profit improved around 15\%.  According to Fig.~\ref{profit}(e), the grid profit improvement is mostly attributable to the lower usage of costly reserve power in the agent-based scenario.  In fact, in this experiment, the grid agent learns to rely on the prosumers' generation for balancing the grid's power rather than using the reserve power which is more expensive.  The use of reserve power is decreased by 10\% in this experiment.

\vspace*{-0.2cm}
\section{Conclusions}~\label{sec:conclusions}
This paper proposes an RL-based distributed energy marketplace framework that enables a real-time, 
demand-dependent, dynamic pricing environment to incentivize prosumers' grid support engagement while improving the economic benefit of both, prosumers and the grid operator.  Simulation results, when implementing the proposed market model, show major economic improvements for the prosumers and the grid (through a reduced reserve power utilization by the grid). 
 \vspace{-0.22in}

\bibliography{bibliography}
\end{document}